\documentclass[a4paper]{article}

\usepackage[top=25mm, bottom=25mm, left=25mm, right=25mm]{geometry}
\input{Qcircuit}

\usepackage{algorithmicx}
\usepackage[ruled]{algorithm}
\usepackage[noset]{algpseudocode}

\newcommand{\euk}{EliminateIdentities}
\newenvironment{alginc}[1][pseudocode]{\medskip\algsetlanguage{#1}\begin{algorithmic}[1]}{\end{algorithmic}\medskip}

\newcommand\ASTART{\bigskip\noindent\begin{minipage}[c]{0.5\linewidth}}

\newcommand\AENDSKIP{\end{minipage}\bigskip}
\newcommand\AEND{\end{minipage}}

\newtheorem{theorem}{Theorem}[section]

\newtheorem{definition}[theorem]{Definition}

\newcommand{\qed}{\nobreak \ifvmode \relax \else
      \ifdim\lastskip<1.5em \hskip-\lastskip
      \hskip1.5em plus0em minus0.5em \fi \nobreak
      \vrule height0.75em width0.5em depth0.25em\fi}

\title{Detection and Elimination of Non-Trivial Reversible Identities}

\author{Ahmed Younes\footnote {ayounes2@yahoo.com or A.Younes@sci.alex.edu.eg}\\
Department of Mathematics and Computer Science\\
Faculty of Science, Alexandria University\\
Alexandria, Egypt}

\begin{document}
\maketitle
\begin{abstract}
Non-Trivial Reversible Identities (NTRIs) are reversible circuits that have equal inputs and outputs. 
NTRIs cannot be detected using optimization algorithms in the literature. 
Existence of NTRIs in a circuit will cause a slow down by increasing the number of gates 
and the quantum cost. NTRIs might arise due to integration 
between two or more optimal reversible circuits. In this paper, an algorithm that detects and removes 
NTRIs in polynomial time will be proposed. Experiments that show the bad effect of NTRIs and the enhancement 
using the proposed algorithm will be presented.

\noindent
Keywords: Reversible Computing; Quantum Cost; Boolean Functions; Circuit Optimization.

\end{abstract}

\section{Introduction}  

Reversible logic \cite{bennett73,fredtoff82} is one of the hot areas of research. It has many applications 
in quantum computation \cite{Gruska99,nc00a}, low-power CMOS \cite{cmos2,cmos1} and many more. 
Synthesis of reversible circuits cannot be done using conventional ways \cite{toffoli80}. 
Synthesis and optimization of Boolean systems on non-standard computers that promise to do computation
more powerfully \cite{simon94} than classical computers, such as
quantum computers, is an essential aim in the exploration of the
benefits that may be gain from such systems.

A function is reversible if it maps an input vector to a unique output vector, 
and vice versa \cite{revlogic}, i.e. we can re-generate the input vector from the output vector (reversibility). 
It is allowed for the inputs to be changed as long as the function maintains reversibility.

A lot of work has been done trying to find an efficient reversible circuit for an arbitrary reversible 
function. Reversible truth table can be seen as a permutation matrix of size $2^n\times2^n$. In one of the research 
directions, it was shown that the process of synthesizing linear reversible circuits can be reduced to 
a row reduction problem of $n \times n$ non-singular matrix \cite{patel}. Standard row reduction methods such as Gaussian 
elimination and LU-decomposition have been proposed \cite{Beth01}. In another research direction, 
search algorithms and template matching tools using reversible gates libraries have been used 
\cite{Dueck,Maslov1,DMMiller2,DMMiller1}. These will work efficiently for small circuits. 
A method is given in \cite{transrules}, where a very useful set
of transformations for Boolean quantum circuits is shown. In this method, extra
auxiliary bits are used in the construction that will increase the hardware cost. In \cite{Younes03b}, 
it was shown that there is a direct correspondence between reversible 
Boolean operations and certain forms of classical logic known as Reed-Muller expansions. 
This arises the possibility of handling the problem of synthesis and optimization of reversible Boolean
logic within the field of Reed-Muller logic. A lot of work has been done trying to find an efficient 
reversible circuit for an arbitrary multi-output Boolean functions by using templates 
\cite{Maslov2,template} and data-structure-based optimization \cite{datastr}. 
A method to generate an optimal 4-bit reversible curcuits has been proposed \cite{optimal4}. In \cite{tool}, 
a very useful set of rules for optimizing reversible circuits and sub-circuits with common-target 
gates has been proposed. Benchmarks for reversible circuits have been established \cite{Benurl2}.

The aim of the paper is to put a highlight on the problem of non-trivial reversible identities (NTRIs) 
that might arise during the process of integrating optimal reversible circuits to do more complex tasks. 
The existence of a NTRI form a bug in the design that will increase the number of gates 
and the quantum cost. The paper proposes an algorithm that detects and removes NITRs in polynomial 
time. The paper is organized as follows. Section 2 gives a short background on reversible gates. 
Section 3 introduces the problem of NTRIs and gives some examples. Section 4 proposes a polynomial time 
algorithm to detect and remove NITRs. Section 5 shows the results of the experiments. The paper ends up
with a conclusion in Section 6.

\section{Background}
\label{sec2}

\begin{center}
\begin{figure}[t]
\begin{center}
\setlength{\unitlength}{3947sp}%
\begingroup\makeatletter\ifx\SetFigFont\undefined%
\gdef\SetFigFont#1#2#3#4#5{%
  \reset@font\fontsize{#1}{#2pt}%
  \fontfamily{#3}\fontseries{#4}\fontshape{#5}%
  \selectfont}%
\fi\endgroup%
\begin{picture}(3169,1438)(5451,-1937)
{\thinlines
\put(6301,-586){\circle*{150}}
}%
{\put(6301,-811){\circle*{150}}}%
{\put(6301,-1261){\circle*{150}}}%
{\put(6301,-1561){\circle{150}}}%
{\put(7658,-1404){\circle{150}}}%
{\put(7661,-1231){\circle*{150}}}%
{\put(8376,-1322){\circle{150}}}%
{\put(8373,-1136){\circle*{150}}}%
{\put(8372,-949){\circle*{150}}}%
{\put(7645,-649){\circle{150}}}%
{\put(6751,-811){\line(-1, 0){900}}}%
{\put(6751,-1261){\line(-1, 0){900}}}%
{\put(6751,-1561){\line(-1, 0){900}}}%
{\put(6301,-511){\line( 0,-1){450}}}%
\put(6270,-1100){\makebox(0,0)[lb]{{$\vdots$}}}

{\put(6301,-1111){\line( 0,-1){525}}}%
{\put(6751,-586){\line(-1, 0){900}}}%
{\put(7424,-1407){\line( 1, 0){463}}}%
{\put(7659,-1162){\line( 0,-1){318}}}%
{\put(7422,-1233){\line( 1, 0){463}}}%
{\put(8145,-1325){\line( 1, 0){463}}}%
{\put(8372,-877){\line( 0,-1){520}}}%
{\put(8137,-1142){\line( 1, 0){463}}}%
{\put(8140,-959){\line( 1, 0){463}}}%
{\put(7418,-650){\line( 1, 0){463}}}%
{\put(7644,-582){\line( 0,-1){140}}}%

\put(5476,-597){\makebox(0,0)[lb]{{$x_{0}$}}}
\put(5476,-814){\makebox(0,0)[lb]{{$x_{1}$}}}
\put(5476,-1267){\makebox(0,0)[lb]{{$x_{n-1}$}}}
\put(5476,-1561){\makebox(0,0)[lb]{{$f_{in}$}}}

\put(6800,-1561){\makebox(0,0)[lb]{{$f_{out}$}}}
\put(6800,-1267){\makebox(0,0)[lb]{{$y_{n-1}$}}}
\put(6800,-814){\makebox(0,0)[lb]{{$y_{1}$}}}
\put(6800,-597){\makebox(0,0)[lb]{{$y_{0}$}}}

\put(7130,-1950){\makebox(0,0)[lb]{{c.Feynman}}}
\put(8131,-1922){\makebox(0,0)[lb]{{d.Toffoli}}}
\put(7350,-939){\makebox(0,0)[lb]{{b.NOT}}}
\put(6000,-1922){\makebox(0,0)[lb]{{a.CNOT}}}
\end{picture}%

\end{center}
\caption{$C^{n}NOT$ gates. The black circle $\bullet $
indicates the control bits, and the symbol $ \oplus $ indicates
the target bit. (a)$C^{n}NOT$ gate with $n$ control bits. (b) $C^{0}NOT$
gate with no control bits. (c)$C^{1}NOT$ gate with one control bit.
(d)$C^{2}NOT$ gate with two control bits.}
\label{figcnot}
\end{figure}
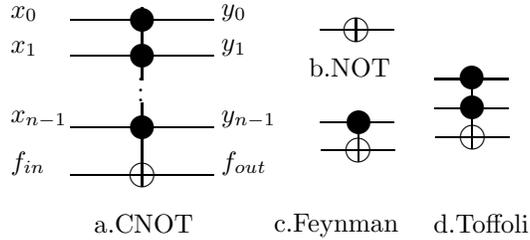
\end{center}

\subsection{Reversible Circuits}

In building a reversible circuit with $n$ variables, an $n \times n$ reversible
circuit will be used. $C^{n}NOT$ gate is the main primitive gate
that will be used in building the circuit since it was shown to be universal for 
reversible computation \cite{toffoli80}. $C^{n}NOT$ gate is defined
as follows:

\begin{definition}($C^{n}NOT$ gate)

$C^{n}NOT$ is a reversible gate denoted as,
\begin{equation} 
C^{n}NOT(x_{n-2}, x_{n-2},\ldots,x_{0};f),
\end{equation}
\noindent
with
$n$ inputs: $x_{n-2}$, $x_{n-2}$,$\ldots,x_{0}$ (known as control
bits) and $f_{in}$ (known as target bit), and $n$ outputs:
$y_{n-2}$, $y_{n-2}$,$\ldots,y_{0}$ and ${f_{out}}$. The operation
of the $C^{n}NOT$ gate is defined as follows,

\begin{equation}
\begin{array}{l}
 y_i  = x_i ,\mathrm{for}\,0 \le i \le n - 2, \\
 f_{out}  = f_{in}  \oplus x_{n - 2} x_{n - 3}  \ldots x_0, \\
 \end{array}
\end{equation}
\noindent
\end{definition}
i.e. the target bit will be flipped if and only if all the control
bits are set to 1. Some special cases of the general $C^{n}NOT$ gate
have their own names, $C^{n}NOT$ gate with no control bits is called
$NOT$ gate as shown in Fig.~\ref{figcnot}-b, where the bit will be
flipped unconditionally. $C^{n}NOT$ gate with one control bit is
called {\it Feynman} gate as shown in Fig.~\ref{figcnot}-c. $C^{n}NOT$ gate
with two control bits is called {\it Toffoli} gate as shown in 
Fig.~\ref{figcnot}-d. For the sake of readability and to keep consistency with the literature, 
$C^{0}NOT$, $C^{1}NOT$, $C^{2}NOT$ and $C^{3}NOT$ will be written for short as $NOT$, $CNOT$, $TOF$ and 
$TOF4$ respectively.

\subsection{Quantum Cost}

{\it Quantum cost} is a term that appears in the literature and is used to refer to
the technological cost of building $C^{n}NOT$ gates. The quantum
cost of a reversible circuit is subject to
optimization as well as the number of $C^{n}NOT$ gates used in the
circuit.  The quantum cost of a $C^{n}NOT$ gate is based primarily on 
the number of bits involved in the gate, i.e. the number of
elementary operations required to build the $C^{n}NOT$ gate
\cite{PhysRevA.52.3457}. The calculation of the quantum cost
for the circuits shown in this paper will be based on the cost table
available in \cite{Benurl2}. The state-of-art shows that both
$NOT(x_i)$ and $CNOT(x_i;f)$ have quantum cost = 1,
$TOF(x_i,x_j;f)$ has a quantum cost = 5, and
$TOF4(x_i,x_j,x_k;f)$ has a quantum cost = 13.

\section{Non-Trivial Reversible Identities}

Circuit identities are usually refer to two or more circuits with the same specification but with different 
designs \cite{tool}. This is usually used to simplify and optimize circuits design \cite{template}. 
In the context of this paper, {\it reversible identities} will refer to circuits that 
output their input without any change, i.e. do nothing. The existence of reversible identities in a circuit 
will increase the number of gates and the quantum cost. Reversible identities can be classified as 
trivial and non-trivial reversible identities.

{\it Trivial reversible identities} are those gates that can be easily detected and removed 
from a reversible circuit. For example, if two adjacent gates are identical then they can be removed 
due to reversibility. If the same concept is applied recursively on a circuit, many gates can be removed. 
For example, consider the circuit $CNOT(b,a)TOF(a,b,c)CNOT(c,b)CNOT(c,b)TOF(a,b,c)\equiv CNOT(b,a)$. 
The two $TOF(a,b,c)$ gates cannot 
be removed together until the two $CNOT(c,b)$ gates are removed. The problem of detection 
and removal of trivial reversible identities are {\it not the main target} of the paper, although 
they will be handled inclusively when dealing with non-trivial reversible identities.

\begin{figure} [t]
\begin{center}
\begin{tabular}
{p{100pt}p{100pt}}
a.
\[
\Qcircuit @C=0.7em @R=0.5em @!R{
&\qw   	&\targ\qwx[1]   	&\targ\qwx[1]   	&\qw   	&\targ\qwx[1]   	&\qw \\
&\qw   	&\ctrl{1}   	&\ctrl{1}   	&\qw   	&\ctrl{1}   	&\qw\\
&\targ\qwx[1]   	&\ctrl{0}   	&\qw\qwx[1]   	&\targ\qwx[1]   	&\ctrl{0}   	&\qw \\
&\ctrl{0}   	&\qw   	&\ctrl{0}   	&\ctrl{0}   	&\qw   	&\qw   	
}
\]
& 
b.
\[
\Qcircuit @C=0.7em @R=0.5em @!R{
&\ctrl{1}   	&\ctrl{1}   	&\ctrl{1}   	&\ctrl{1}   	&\ctrl{1}   	&\ctrl{1}   	&\qw \\
&\targ\qwx[1]   	&\ctrl{1}   	&\targ\qwx[1]   	&\ctrl{1}   	&\targ\qwx[1]   	&\ctrl{1}   	&\qw\\
&\ctrl{1}   	&\targ\qwx[1]   	&\ctrl{1}   	&\targ\qwx[1]   	&\ctrl{1}   	&\targ\qwx[1]   	&\qw\\
&\ctrl{0}   	&\ctrl{0}   	&\ctrl{0}   	&\ctrl{0}   	&\ctrl{0}   	&\ctrl{0}   	&\qw   	
}
\]
\\
\end{tabular}
\end{center}
\caption{Non-trivial Reversible Identities.}
\label{ABC}
\end{figure}
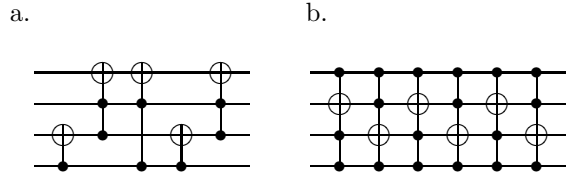

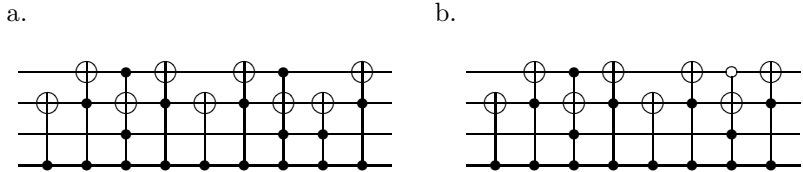
\begin{figure} [t]
\begin{center}
\begin{tabular}
{p{150pt}p{150pt}}
a.
\[
\Qcircuit @C=0.7em @R=0.5em @!R{
   	&\qw   	&\targ\qwx[1]   	&\ctrl{1}   	&\targ\qwx[1]   	&\qw   	&\targ\qwx[1]   	&\ctrl{1}   	&\qw   	&\targ\qwx[1]   	&\qw   	  	\\
   	&\targ\qwx[1]   	&\ctrl{1}   	&\targ\qwx[1]   	&\ctrl{1}   	&\targ\qwx[1]   	&\ctrl{1}   	&\targ\qwx[1]   	&\targ\qwx[1]   	&\ctrl{1}   	&\qw   	   	\\
   	&\qw\qwx[1]   	&\qw\qwx[1]   	&\ctrl{1}   	&\qw\qwx[1]   	&\qw\qwx[1]   	&\qw\qwx[1]   	&\ctrl{1}   	&\ctrl{1}   	&\qw\qwx[1]   	&\qw   	   	\\
   	&\ctrl{0}   	&\ctrl{0}   	&\ctrl{0}   	&\ctrl{0}   	&\ctrl{0}   	&\ctrl{0}   	&\ctrl{0}   	&\ctrl{0}   	&\ctrl{0}   	&\qw   	   	
}
\]
& 
b.
\[
\Qcircuit @C=0.7em @R=0.5em @!R{
&\qw   	&\targ\qwx[1]   	&\ctrl{1}   	&\targ\qwx[1]   	&\qw   	&\targ\qwx[1]   	&\ctrlo{1}   	&\targ\qwx[1]   	&\qw   	   	\\
&\targ\qwx[1]   	&\ctrl{1}   	&\targ\qwx[1]   	&\ctrl{1}   	&\targ\qwx[1]   	&\ctrl{1}   	&\targ\qwx[1]   	&\ctrl{1}   	&\qw   	   	\\
&\qw\qwx[1]   	&\qw\qwx[1]   	&\ctrl{1}   	&\qw\qwx[1]   	&\qw\qwx[1]   	&\qw\qwx[1]   	&\ctrl{1}   	&\qw\qwx[1]   	&\qw   	   	\\
&\ctrl{0}   	&\ctrl{0}   	&\ctrl{0}   	&\ctrl{0}   	&\ctrl{0}   	&\ctrl{0}   	&\ctrl{0}   	&\ctrl{0}   	&\qw     	
}
\]
\\
\end{tabular}
\end{center}
\caption{A NTRI before, part-a, and after, part-b, optimization.}
\label{figIDopt}
\end{figure}

{\it Non-trivial reversible identities (NTRIs)} are reversible identities that cannot be detected and removed 
using any know optimization algorithm, for example, \cite{tool,template,datastr}. Fig.~\ref{ABC}
shows two examples of NTRIs. Applying any known optimization algorithm will report that these 
designs cannot be optimized further. Fig.~\ref{figIDopt}-a shows another example of 
NTRI that is slightly optimized using \cite{tool} by decreasing 
the number of gates by one as shown in Fig.~\ref{figIDopt}-b. {\it NTRI must be removed 
completely from the circuit as they form a bug in the design}.

It is important here to differentiate between synthesis and optimization of reversible circuits 
to show a situation where NTRI might arise as a bug. Synthesis of a reversible circuit is the process 
of constructing, from scratch, the best design for a given specification. 
Synthesis process always includes optimization. Optimization of a reversible circuit is the process 
of enhancing an existing design for a given specification by decreasing the number of gates 
and/or the quantum cost for a given circuit. NTRI is not likely to occur during the synthesizing 
process, while NTRI might arise as a problem during the optimization process.

\begin{figure} [t]
\begin{center}
\begin{tabular}
{p{330pt}}

\[
\Qcircuit @C=0.7em @R=0.5em @!R{
&a.   	&   	&   	&   \downarrow	&   	&   	&  \\
&\ctrl{1}   	&\ctrl{1}   	&\targ\qwx[1]   	&\ctrl{1}   	&\ctrl{1}   	&\ctrl{1}   	&\qw \\
&\qw\qwx[1]   	&\ctrl{1}   	&\qw\qwx[1]   	&\targ\qwx[1]   	&\qw\qwx[1]   	&\targ\qwx[1]   	&\qw \\
&\targ\qwx[1]   	&\qw\qwx[1]   	&\ctrl{1}   	&\ctrl{1}   	&\targ\qwx[1]   	&\ctrl{1}   	&\qw\\
&\ctrl{0}   	&\targ   	&\ctrl{0}   	&\ctrl{0}   	&\ctrl{0}   	&\ctrl{0}   	&\qw    	
}
\]

\[
\Qcircuit @C=0.7em @R=0.5em @!R{
& b.   	&   	&   	&   	&   	&   	&   	&   	&   	&   	&   	&   	&   	&   	&  	&   	&   	& \\
&\ctrl{1}   	&\ctrl{1}   	&\ctrl{1}   	&\ctrl{1}   	&\ctrl{1}   	&\qw   	&\qw   	&\qw   	&\qw   	&\ctrl{1}   	&\qw   	&\targ\qwx[1]   	&\qw   	&\targ\qwx[1]   	&\qw   	&\ctrl{1}   	&\targ   	&\qw\\
&\qw\qwx[1]   	&\targ\qwx[1]   	&\qw\qwx[1]   	&\targ\qwx[1]   	&\qw\qwx[1]   	&\ctrl{1}   	&\ctrl{1}   	&\qw   	&\qw   	&\targ\qwx[1]   	&\targ\qwx[1]   	&\qw\qwx[1]   	&\ctrl{1}   	&\ctrl{1}   	&\targ   	&\qw\qwx[1]   	&\qw   	&\qw   	\\
&\targ\qwx[1]   	&\ctrl{1}   	&\targ\qwx[1]   	&\ctrl{1}   	&\targ\qwx[1]   	&\ctrl{1}   	&\targ   	&\targ   	&\ctrl{1}   	&\ctrl{0}   	&\ctrl{1}   	&\qw\qwx[1]   	&\targ   	&\ctrl{1}   	&\qw   	&\targ\qwx[1]   	&\qw   	&\qw\\
&\ctrl{0}   	&\ctrl{0}   	&\ctrl{0}   	&\ctrl{0}   	&\ctrl{0}   	&\targ   	&\qw   	&\qw   	&\targ   	&\qw   	&\ctrl{0}   	&\ctrl{0}   	&\qw   	&\ctrl{0}   	&\qw   	&\ctrl{0}   	&\qw   	&\qw	\\   	
&    			&    			&    			&    	&    \uparrow			&    		&    	&   	&    		&   	&   			&   			&   	&  				&   	&   			& 		&
}
\]
\\
\end{tabular}
\end{center}
\caption{Two Reversible circuits are required to 
be integrated such that circuit in part-b should be added to the rear of circuit in part-a.}
\label{fig2parts}
\end{figure}
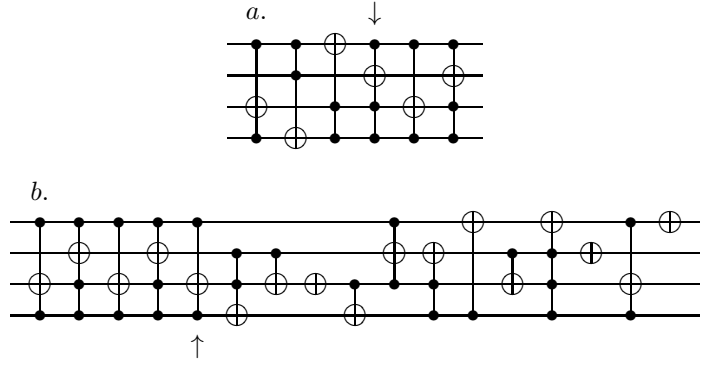

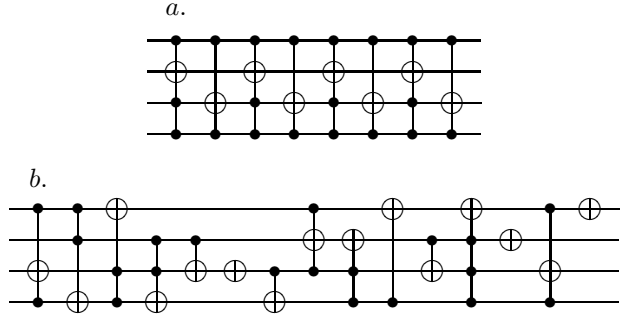
\begin{figure} [t]
\begin{center}
\begin{tabular}
{p{330pt}}

\[
\Qcircuit @C=0.7em @R=0.5em @!R{
&a.   	&  &   	&   	&   	&   	&   	&   	&  \\
&\ctrl{1}   	&\ctrl{1}   	&\ctrl{1}   	&\ctrl{1}   	&\ctrl{1}   	&\ctrl{1}   	&\ctrl{1}   	&\ctrl{1}   	&\qw\\
&\targ\qwx[1]   	&\qw\qwx[1]   	&\targ\qwx[1]   	&\qw\qwx[1]   	&\targ\qwx[1]   	&\qw\qwx[1]   	&\targ\qwx[1]   	&\qw\qwx[1]   	&\qw\\
&\ctrl{1}   	&\targ\qwx[1]   	&\ctrl{1}   	&\targ\qwx[1]   	&\ctrl{1}   	&\targ\qwx[1]   	&\ctrl{1}   	&\targ\qwx[1]   	&\qw\\
&\ctrl{0}   	&\ctrl{0}   	&\ctrl{0}   	&\ctrl{0}   	&\ctrl{0}   	&\ctrl{0}   	&\ctrl{0}   	&\ctrl{0}   	&\qw 	
}
\]

\[
\Qcircuit @C=0.7em @R=0.5em @!R{
&b.   	&   	&   	&   	&   	&   	&   	&   	&   	&   	&   	&   	&  	&   	&   	&  \\
&\ctrl{1}   	&\ctrl{1}   	&\targ\qwx[1]   	&\qw   	&\qw   	&\qw   	&\qw   	&\ctrl{1}   	&\qw   	&\targ\qwx[1]   	&\qw   	&\targ\qwx[1]   	&\qw   	&\ctrl{1}   	&\targ   	&\qw\\
&\qw\qwx[1]   	&\ctrl{1}   	&\qw\qwx[1]   	&\ctrl{1}   	&\ctrl{1}   	&\qw   	&\qw   	&\targ\qwx[1]   	&\targ\qwx[1]   	&\qw\qwx[1]   	&\ctrl{1}   	&\ctrl{1}   	&\targ   	&\qw\qwx[1]   	&\qw   	&\qw\\
&\targ\qwx[1]   	&\qw\qwx[1]   	&\ctrl{1}   	&\ctrl{1}   	&\targ   	&\targ   	&\ctrl{1}   	&\ctrl{0}   	&\ctrl{1}   	&\qw\qwx[1]   	&\targ   	&\ctrl{1}   	&\qw   	&\targ\qwx[1]   	&\qw   	&\qw\\
&\ctrl{0}   	&\targ   	&\ctrl{0}   	&\targ   	&\qw   	&\qw   	&\targ   	&\qw   	&\ctrl{0}   	&\ctrl{0}   	&\qw   	&\ctrl{0}   	&\qw   	&\ctrl{0}   	&\qw   	&\qw 	
}
\]
\\
\end{tabular}
\end{center}
\caption{Circuit in part-a shows the discovered NTRI (the bug) and circuit in part-b shows the final circuit after 
elimination of the NTRI.}
\label{figcirID}
\end{figure}

To Design a reversible circuit that does a complex task, it will not be practical to do 
the synthesis process of that circuit from scratch. Instead, the complex task should be broken down 
to a set of simple modules, design the optimal form for each module then integrate the optimal designs 
in a complete design that perform the required task. NTRI might arise during the process of integration. 
For example, consider the two reversible circuits shown in Fig.~\ref{fig2parts}. These circuits cannot be optimized 
further using any known optimization algorithm \cite{tool}. 
Consider that these two circuits should be integrated such that circuit in  Fig.~\ref{fig2parts}-b 
should be added to the rear of circuit in Fig.~\ref{fig2parts}-a. The integrated circuit contains 
23 gates with quantum cost = 125. No further optimization can be done on the integrated circuit 
although it is bugged by NRTI in the middle of the integrated circuit. Fig.~\ref{figcirID}-a shows 
the NRTI and Fig.~\ref{figcirID}-b shows the circuit with 15 gates and 
quantum cost = 53 after removing the NTRI. Arrows in Fig.~\ref{fig2parts}-a and Fig.~\ref{fig2parts}-b 
mark the start and the end of the NTRI respectively. In the next section, an algorithm 
to remove NTRIs in polynomial time will be proposed. It is important to notice that this algorithm 
is not a substitution for the current optimization algorithms in the literature, this algorithm 
is recommended to be used as a module with any optimization algorithm.

\section{Detection and Elimination of NTRIs}
\subsection{Data Structures}
Consider a finite set $A=\{0,1,...,N-1\}$ and a bijection $\sigma :A \to A$,
then $\sigma$ can be written as,

\begin{equation}
\sigma  = \left( {\begin{array}{*{20}c}
   0 & 1 & 2 & {...} & {N - 1}  \\
   {\sigma (0)} & {\sigma (1)} & {\sigma (2)} & {...} & {\sigma (N - 1)}  \\
\end{array}} \right),
\end{equation}

\noindent i.e. $\sigma$ is a permutation of $A$. Let $A$ be an ordered set, then the top row can be 
eliminated and $\sigma$ can be written as,

\begin{equation}
\left[ {\sigma (0)} , {\sigma (1)} , {\sigma (2)} , {...} , {\sigma (N - 1)}\right].
\label{specs}
\end{equation}

Any reversible circuit with $n$ inputs can be considered as a permutation $\sigma$ and Eqn.(\ref{specs}) is 
the {\it specification} of this reversible circuit such that $N=2^n$.

In order to detect reversible identities, it is required to store the output of the circuit 
after applying every gate. Two data structures are used in the proposed algorithm, {\it Circuit} and 
{\it CurrentSpecs}. The first data structure, {\it Circuit}, is a one dimensional array of size $m$, where $m$ is the number 
of gates in the reversible circuit. {\it Circuit} stores the gates of the reversible circuit such that,
{\it Circuit[1]} contains the first gate and {\it Circuit[m]} contains the last gate.

The second data structure, {\it CurrentSpecs}, is used to store the output specification after 
applying every gate in the reversible circuit. {\it CurrentSpecs} is a two dimensional array 
of size $N \times m$.

\subsection{The Algorithm}

The outline of the proposed algorithm is as follows: 
Run the circuit by applying one gate at a time. Store the specification after applying each gate, 
and compare the current specification at position $i$ with all the previously stored specifications. 
If specification at position $i$ is equal to specification at position $j$, where $j<i$, then remove 
gates from point $j$ to point $i$. Repeat the detection until no more reversible identities are found.

\begin{algorithm}[H]
\caption{Eliminate NTRIs algorithm}\label{EIdent}
\begin{alginc}
\Procedure{\euk}{$Circuit$}
\State $Finish\gets false$

\While{$Finish=false$} \label{whileflag}
	\State $Finish\gets true$
	\State $CurrentCircuit \gets nil$
	\State $m\gets SIZE(Circuit)$

	\For{$i\gets 1, m$}
			\State $CurrentCircuit\gets Circuit[i]\cup CurrentCircuit$ 
			\State $CurrentSpecs[i]\gets SPECS(CurrentCircuit)$
			\For{$j\gets 1, i-1$}
				\If{$CurrentSpecs[j]= CurrentSpecs[i]$}
					\State $REMOVEGATES(Circuit,j,i)$				
					\State $Finish\gets false$
					\State $j \gets i-1$		\Comment{Finish inner loop early}
					\State $i \gets m$			\Comment{Finish outer loop early}
				\EndIf
            \EndFor	
     \EndFor
\EndWhile		
\EndProcedure
\end{alginc}
\end{algorithm}

The correctness of the algorithm is proved as follows. The {\bf while loop} from Line:3 to Line:19 will 
repeat until no more reversible identities exist in {\it Circuit}. The {\bf for loop} from Line:7 to Line:18 traces {\it Circuit} one gate at a time
and stores the circuit from gate 0 to gate $i$ in {\it CurrentCircuit}. {\it CurrentSpecs[i]} stores 
the specification for the circuit at point $i$. The {\bf for loop} from Line:10 to Line:17 checks 
if there exist any reversible identity by comparing specification of point $i$ with specification of point $j$ 
such that $1\le j<i$. At Line:11, if an identity is found, then gates from point $j$ to point $i$ is 
removed from {\it Circuit}, and the algorithm starts all over again looking for more reversible identities.

The best case running time exists when no identities exist in {\it Circuit} 
where the {\bf while loop} from Line:3 to Line:19 will run once so the algorithm has $\Theta \left( {m^2 } \right)$. 
The worst case running time exists when {\it Circuit} contains gates such that every two 
adjacent gates are identical where the {\bf while loop} will repeat $m/2$ times, so 
the algorithm has $\Theta \left( {m^3 } \right)$. The proposed algorithm can detect and eliminate identities from the reversible circuit in polynomial time.

\section{Experimental Results }

Based on literature review, no method proposed the problem of NTRIs so far. To test the effect of NTRIs on 
the optimization of reversible circuits. Two software's have been developed.
The first generates random reversible circuits with NTRIs and the second generates random NTRIs. 
Experiments are based on 4-bits reversible circuits as a prototype.

The first experiment is done by inserting a random NTRI in a randomly chosen insertion point in the 
middle of optimal 4-bits reversible circuits \cite{optimal4}. Then the latest optimization method \cite{tool} 
using \cite{toolsw} is applied to test the effect of the NTRI 
on the final number of gates and the quantum cost of the circuit. 
Table \ref{tab1} shows the results of the experiment where NTRIs affect the number of gates 
and the quantum cost. Applying Algorithm~\ref{EIdent} before \cite{tool} gives the optimal results. 
Circuits used in the experiment are shown in Appendix 1.


\begin{table}[t]
\begin{center}
{\footnotesize
\begin{tabular}
{|p{45pt}|p{40pt}|p{35pt}|p{30pt}|p{30pt}|p{30pt}|p{35pt}|}
\hline
Benchmark& 
Optimal Circuit\cite{optimal4}& 
Random Identity& 
Insertion Pt.& 
Optimal (g,c)\cite{optimal4}& 
Bugged \par (g,c) & 
Optimized (g,c)\cite{tool} \\
\hline
4{\_}49& 
APP1.1.a \par & 
APP1.1.b \par & 
6& 
(12,32)& 
(19,61)& 
(19,61) \\
\hline
4bit-7-8& 
APP1.2.a \par & 
APP1.2.b \par & 
5& 
(7,19)& 
(14,40)& 
(12,34) \\
\hline
decode42& 
APP1.3.a \par & 
APP1.3.b \par & 
4& 
(10,30)& 
(16,52)& 
(15,51) \\
\hline
hwb4& 
APP1.4.a \par & 
APP1.4.b \par & 
7& 
(11,39)& 
(16,64)& 
(16,62) \\
\hline
imark& 
APP1.5.a \par & 
APP1.5.b \par & 
5& 
(7,19)& 
(17,43)& 
(11,37) \\
\hline
mperk& 
APP1.6.a  \par & 
APP1.6.b \par & 
4& 
(9,15)& 
(22,52)& 
(22,52) \\
\hline
oc5& 
APP1.7.a \par & 
APP1.7.b \par & 
2& 
(11,39)& 
(23,65)& 
(16,52) \\
\hline
oc6& 
APP1.8.a \par & 
APP1.8.b \par & 
11& 
(12,60)& 
(20,74)& 
(20,74) \\
\hline
oc7& 
APP1.9.a \par & 
APP1.9.a \par & 
13& 
(13,41)& 
(29,219)& 
(28,207) \\
\hline
oc8& 
APP1.10.a \par & 
APP1.10.b \par & 
9& 
(11,47)& 
(25,197)& 
(15,79) \\
\hline
primes4& 
APP1.11.a \par & 
APP1.11.b \par & 
4& 
(10,42)& 
(18,98)& 
(13,77) \\
\hline
rd32& 
APP1.12.a \par & 
APP1.12.b \par & 
3& 
(4,8)& 
(10,54)& 
(8,46) \\
\hline
shift4& 
APP1.13.a \par & 
APP1.13.b \par & 
4& 
(4,18)& 
(20,146)& 
(13,101) \\
\hline
\end{tabular}

}
\end{center}
\caption{Optimal 4-bits reversible circuits from \cite{optimal4} bugged by random NTRIs and optimized 
using \cite{tool}, where the pair $(g,c)$ represents the number of gates $(g)$ and the quantum cost $(c)$. }
\label{tab1}
\end{table}

The second experiment is used to test Algorithm~\ref{EIdent} by generating random reversible circuits and checking 
the number of gates and the quantum cost in the following cases: (1) before applying any method, (2) after applying \cite{tool}, 
(3) after applying the proposed method then \cite{tool}. Table \ref{tab2} shows the results of the experiment with 
a better result using the proposed algorithm. Circuits used in the experiment are shown in Appendix 2.

\begin{table}[t]
\begin{center}
{\footnotesize
\begin{tabular}
{|p{80pt}|p{40pt}|p{40pt}|p{40pt}|p{40pt}|}
\hline
Specification& 
Random Circuit& 
Original (g,c)& 
Optimized (g,c) \cite{tool}& 
Optimized + NTRIs Removal (g,c)  \\
\hline
[12,7,2,5,0,15,14,11,  6,3,10,1,8,9,4,13]& 
APP2.1 \par & 
(21,113)& 
(17,103)& 
(10,30) \\
\hline
[7,14,9,6,11,0,13,2,    5,15,10,12,1,4,3,8]& 
APP2.2 \par & 
(30,210)& 
(30,204)& 
(18,102) \\
\hline
[10,15,0,7,14,9,6,1,    13,12,5,3,11,8,4,2]& 
APP2.3 & 
(23,103)& 
(23,101)& 
(13,43) \\
\hline
[12,9,11,14,6,7,8,10,    2,3,4,5,15,13,0,1]& 
APP2.4 \par & 
(22,90)& 
(22,90)& 
(9,36) \\
\hline
[0,1,15,8,4,5,9,14,    11,12,7,6,3,13,10,2]& 
APP2.5 \par & 
(23,137)& 
(19,105)& 
(10,50) \\
\hline
[3,0,1,6,7,2,5,4,        11,8,9,14,15,10,13,12]& 
APP2.6 \par & 
(25,133)& 
(19,99)& 
(6,14) \\
\hline
[6,11,5,4,2,0,1,      15,14,3,12,8,7,9,13,10]& 
APP2.7 \par & 
(21,137)& 
(20,132)& 
(15,59) \\
\hline
[12,15,5,8,3,2,1,10,      7,14,13,6,11,0,9,4]& 
APP2.8 \par & 
(23,125)& 
(23,125)& 
(15,53) \\
\hline
[0,1,6,5,7,8,15,2,          14,13,12,3,11,4,9,10]& 
APP2.9 \par & 
(17,65)& 
(16,64)& 
(11,47) \\
\hline
[0,10,2,15,8,9,4,1,         6,5,14,3,12,13,11,7]& 
APP2.10 \par & 
(20,80)& 
(19,75)& 
(13,57) \\
\hline
[8,9,10,2,4,7,6,5,     0,15,13,3,12,14,1,11]& 
APP2.11 \par & 
(21,93)& 
(21,93)& 
(12,80) \\
\hline
[6,15,0,1,9,2,7,4,      11,10,5,12,3,14,13,8]& 
APP2.12 \par & 
(29,73)& 
(29,73)& 
(17,53) \\
\hline
[9,3,10,11,12,13,1,7,     0,8,14,2,15,4,5,6]& 
APP2.13 \par & 
(25,81)& 
(17,69)& 
(12,52) \\
\hline
\end{tabular}
}
\end{center}
\caption{Random reversible circuits with NTRIs optimized using \cite{tool} alone and optimized 
using a hybrid system from the proposed algorithm and \cite{tool}. where the pair $(g,c)$ represents the number of gates $(g)$ and the quantum cost $(c)$.}
\label{tab2}
\end{table}

\section{Conclusion}

Non-trivial reversible identities (NTRIs) are bugs that might arise in integrated reversible circuits 
to cause a slow down, increase in the number of gates and the quantum cost of the integrated 
circuits. A polynomial time algorithm is proposed to detect and eliminate NTRIs. The proposed algorithm 
is not a substitution of any optimization algorithm in the literature. The paper recommends 
the proposed algorithm to be used together with any optimization algorithm to handle the problem of NTRIs.

\appendix
{\footnotesize
\section{Appendix 1}

Circuits used in the first experiment where "\#" shows the insertion point of the NTRI bug.

\subsection*{APP1.1.a}
NOT(a) CNOT(c, a) CNOT(a, d) TOF(a, b, d) CNOT(d, a) \# TOF(c, d, b) TOF(a, d, c) TOF(b, c, a) TOF(a, b, d) NOT(a) CNOT(d, b) CNOT(d, c)

\subsection*{APP1.1.b}
TOF(b, d, c) CNOT(c, a) CNOT(d, c) CNOT(c, a) TOF4(a, b, d, c) CNOT(d, a) TOF4(a, b, d, c) CNOT(d, c)

\subsection*{APP1.2.a}
CNOT(d, b) CNOT(d, a) CNOT(c, d) TOF4(a, b, d, c) \# CNOT(c, d) CNOT(d, b) CNOT(d, a)          
\subsection*{APP1.2.b}
CNOT(d, b) TOF(a, d, c) CNOT(c, a) TOF(a, d, b) CNOT(d, b) CNOT(c, a) TOF(a, d, c) TOF(c, d, b) 
\subsection*{APP1.3.a}
CNOT(c, b) CNOT(d, a) CNOT(c, a) \# TOF(a, d, b) CNOT(b, c) TOF4(a, b, c, d) TOF(b, d, c) CNOT(c, a) CNOT(a, b) NOT(a)    
\subsection*{APP1.3.b}
TOF(b, d, a) CNOT(d, b) TOF(c, d, b) CNOT(d, b) TOF(c, d, a) TOF(b, d, a) TOF(c, d, b) 
\subsection*{APP1.4.a}
CNOT(b, d) CNOT(d, a) CNOT(a, c) TOF4(b, c, d, a) CNOT(d, b) CNOT(c, d) TOF(a, c, b) \# TOF4(b, c, d, a) CNOT(d, c) CNOT(a, c) CNOT(b, d)  
\subsection*{APP1.4.b}
TOF(c, d, b) TOF(a, d, c) TOF(a, d, b) TOF(c, d, b) TOF(a, d, c) 
\subsection*{APP1.5.a}
TOF(c, d, a) TOF(a, b, d) CNOT(d, c) CNOT(b, c) \# CNOT(d, a) TOF(a, c, b) NOT(c) 
\subsection*{APP1.5.b}
TOF(a, d, b) CNOT(c, a) CNOT(d, a) CNOT(c, a) TOF(c, d, b) CNOT(d, a) CNOT(d, b) TOF(c, d, b) CNOT(d, b) TOF(a, d, b) 
\subsection*{APP1.6.a}
NOT(c) CNOT(d, c) TOF(c, d, b) \# TOF(a, c, d) CNOT(b, a) CNOT(d, a) CNOT(c, a) CNOT(a, b) CNOT(b, c) 
\subsection*{APP1.6.b}
CNOT(d, b) CNOT(d, a) CNOT(c, a) TOF(a, d, b) CNOT(d, b) CNOT(c, a) TOF(a, d, b) TOF(a, d, b) TOF(c, d, a) CNOT(d, a) CNOT(d, b) TOF(a, d, b) TOF(c, d, a) 
\subsection*{APP1.7.a}
TOF(b, d, c) \# TOF(c, d, b) TOF(a, b, c) NOT(a) CNOT(d, b) CNOT(a, c) TOF(b, c, d) CNOT(a, b) CNOT(c, a) CNOT(a, c) TOF4(a, b, d, c) 
\subsection*{APP1.7.b}
CNOT(d, b) CNOT(c, a) CNOT(d, a) CNOT(c, a) CNOT(d, b) TOF(c, d, a) CNOT(d, a) TOF(c, d, b) CNOT(d, b) TOF(c, d, b) TOF(c, d, a) CNOT(d, b) 
\subsection*{APP1.8.a}
TOF4(b, c, d, a) TOF4(a, c, d, b) CNOT(d, c) TOF(b, c, d) TOF(c, d, a) TOF4(a, b, d, c) CNOT(b, a) NOT(a) CNOT(c, b) CNOT(d, c) CNOT(a, d) \# TOF(b, d, c) 
\subsection*{APP1.8.b}
TOF(c, d, a) CNOT(c, a) CNOT(d, b) TOF(c, d, a) CNOT(d, a) CNOT(d, a) CNOT(c, a) CNOT(d, b) 
\subsection*{APP1.9.a}
TOF(b, d, c) TOF(a, b, d) CNOT(b, a) TOF4(a, c, d, b) CNOT(c, b) CNOT(d, c) TOF(a, c, d) NOT(b) NOT(d) CNOT(b, c) TOF(b, d, a) TOF(a, c, d) \# CNOT(c, a)    
\subsection*{APP1.9.a}
TOF4(a, b, d, c) TOF(a, d, b) TOF4(a, b, d, c) TOF4(a, c, d, b) TOF4(a, b, d, c) TOF(a, d, b) TOF4(a, d, c, b) TOF4(a, b, d, c) TOF4(a, d, c, b) TOF(a, d, b) TOF4(a, c, d, b) TOF4(a, b, d, c) TOF(a, d, b) TOF4(a, c, d, b) TOF4(a, b, d, c) TOF4(a, c, d, b) 
\subsection*{APP1.10.a}
CNOT(d, a) TOF(b, c, a) TOF(c, d, b) TOF4(a, b, d, c) TOF(a, b, d) TOF(a, d, b) NOT(a) NOT(b) TOF(b, d, a) \# CNOT(a, d) TOF(b, c, d) 
\subsection*{APP1.10.b}
TOF(a, d, b) TOF(a, d, b) TOF4(a, c, d, b) TOF4(a, b, d, c) TOF4(a, d, c, b) TOF(a, d, b) TOF4(a, c, d, b) TOF4(a, b, d, c) TOF4(a, b, d, c) TOF4(a, c, d, b) TOF(a, d, b) TOF4(a, c, d, b) TOF4(a, b, d, c) TOF4(a, c, d, b) 
\subsection*{APP1.11.a}
CNOT(d, c) CNOT(c, a) CNOT(b, c) \# NOT(b) TOF(b, c, d) TOF4(a, b, d, c) TOF(a, c, b) NOT(a) TOF4(a, c, d, b) CNOT(b, a)     
\subsection*{APP1.11.b}
TOF(a, c, b) TOF4(a, c, d, b) TOF(a, d, b) TOF(b, d, a) TOF(b, d, a) TOF(a, d, b) TOF4(a, c, d, b) TOF(a, c, b) 
\subsection*{APP1.12.a}
TOF(a, b, d) CNOT(a, b) \# TOF(b, c, d) CNOT(b, c) 
\subsection*{APP1.12.b}
TOF(c, d, b) TOF4(a, c, d, b) TOF(a, d, b) TOF4(a, c, d, b) TOF(c, d, b) TOF(a, d, b) 
\subsection*{APP1.13.a}
TOF4(a, b, c, d) TOF(a, b, c) CNOT(a, b) \# NOT(a)      
\subsection*{APP1.13.b}
TOF(a, d, b) TOF(b, d, c) TOF(a, d, b) TOF(a, d, b) TOF4(a, c, d, b) TOF4(a, b, d, c) TOF(b, d, c) TOF4(a, c, d, b) TOF(b, d, c) TOF(a, d, b) TOF(b, d, c) TOF(a, d, b) TOF4(a, c, d, b) TOF(a, d, b) TOF4(a, c, d, b) TOF4(a, b, d, c)

\section{Appendix 2}

Circuits used in the second experiment where [...] shows the NTRI discovered by the proposed algorithm.
\subsection*{APP2.1}
CNOT(b, c) NOT(b) TOF4(a, b, c, d) CNOT(a, b) [TOF(a, d, b) TOF(b, d, c) TOF4(a, c, d, b) TOF4(a, b, d, c) TOF(b, d, c) TOF4(a, c, d, b) TOF(b, d, c) TOF(a, d, b) TOF(b, d, c) TOF4(a, b, d, c) ]CNOT(a, d) TOF(a, d, b) NOT(b) TOF(a, b, c) NOT(c) TOF(c, d, b) CNOT(c, d) 
\subsection*{APP2.2}
TOF4(a, c, d, b) CNOT(c, a) TOF4(a, c, d, b) NOT(c) TOF(b, d, c) TOF(a, d, c) [TOF4(a, c, d, b) TOF(a, d, b) TOF4(a, b, d, c) TOF(a, d, b) TOF4(a, b, d, c) TOF(a, d, b) TOF(a, d, c) TOF4(a, c, d, b) TOF4(a, b, d, c) TOF(a, d, b) TOF(a, d, c) TOF4(a, b, d, c) ]TOF(a, b, d) TOF(a, d, b) TOF4(b, c, d, a) CNOT(a, b) CNOT(c, a) CNOT(c, d) TOF(a, d, b) TOF4(a, b, c, d) CNOT(b, c) TOF(a, d, c) TOF4(a, b, c, d) CNOT(b, c) 
\subsection*{APP2.3}
TOF(a, d, c) TOF4(b, c, d, a) CNOT(d, a) CNOT(c, b) NOT(b) [CNOT(d, b) TOF(b, d, a) TOF(c, d, b) TOF(b, d, a) TOF4(a, c, d, b) CNOT(d, b) TOF(b, d, a) TOF(c, d, b) TOF(b, d, a) TOF4(a, c, d, b) TOF(c, d, b) ]CNOT(d, b) CNOT(b, d) CNOT(c, d) TOF4(a, c, d, b) CNOT(a, c) TOF(c, d, b) CNOT(a, b) 
\subsection*{APP2.4}
NOT(c) NOT(d) NOT(b) CNOT(a, d) [CNOT(d, b) TOF(b, d, a) TOF4(a, c, d, b) TOF(b, d, a) CNOT(d, b) TOF(b, d, a) TOF4(a, c, d, b) TOF(c, d, b) TOF(b, d, a) ]CNOT(b, c) CNOT(a, d) TOF(b, c, d) TOF(a, d, b) CNOT(b, c) TOF(c, d, b) TOF4(a, c, d, b) CNOT(b, c) TOF(b, d, a) 
\subsection*{APP2.5}
TOF(c, d, a) TOF(a, b, c) TOF(a, d, b) CNOT(b, c) [TOF4(a, c, d, b) TOF(c, d, b) TOF(b, d, a) TOF4(a, c, d, b) CNOT(d, b) TOF(c, d, b) TOF(b, d, a) TOF(c, d, b) TOF(b, d, a) TOF4(a, c, d, b) CNOT(d, b) TOF(b, d, a) TOF4(a, c, d, b) ]TOF4(a, c, d, b) TOF4(b, c, d, a) CNOT(b, d) CNOT(d, b) CNOT(d, a) TOF(c, d, b)  
\subsection*{APP2.6}
TOF4(a, c, d, b) CNOT(c, b) TOF4(a, c, d, b) CNOT(a, b) [TOF4(a, c, d, b) TOF(b, d, a) TOF4(a, c, d, b) TOF(b, d, a) CNOT(d, b) TOF(b, d, a) TOF4(a, c, d, b) TOF(b, d, a) CNOT(d, b) TOF4(a, c, d, b) TOF(c, d, b) TOF(b, d, a) TOF(c, d, b) TOF(b, d, a) ]CNOT(c, b) NOT(a) NOT(b) CNOT(c, b) TOF(a, c, b) CNOT(b, c) TOF(a, b, c) 
\subsection*{APP2.7}
CNOT(a, d) TOF(a, c, b) CNOT(c, d) TOF(a, b, c) CNOT(c, d) CNOT(b, a) TOF4(a, b, c, d) TOF(a, c, d) TOF(c, d, a) NOT(b) CNOT(b, c) [TOF4(a, c, d, b) TOF4(a, b, d, c) TOF4(a, c, d, b) TOF4(a, b, d, c) TOF4(a, c, d, b) TOF4(a, b, d, c) ]CNOT(a, c) TOF4(a, b, c, d) TOF(b, d, a) CNOT(d, a) 
\subsection*{APP2.8}
TOF(a, d, c) TOF(a, b, d) TOF(c, d, a) [TOF4(a, c, d, b) TOF(a, d, c) TOF4(a, c, d, b) TOF(a, d, c) TOF4(a, c, d, b) TOF(a, d, c) TOF4(a, c, d, b) TOF(a, d, c) ]TOF(b, c, d) CNOT(b, c) NOT(c) CNOT(c, d) TOF(a, c, b) TOF(c, d, b) CNOT(d, a) CNOT(b, c) TOF4(b, c, d, a) NOT(b) TOF(a, d, c) NOT(a) 
\subsection*{APP2.9}
TOF(a, c, d) TOF(a, d, b) CNOT(d, b) CNOT(c, a) TOF(a, c, b) TOF(a, b, c) TOF(b, d, c) TOF(a, c, b) TOF4(a, b, c, d) [CNOT(d, c) TOF(b, c, a) TOF(b, d, a) CNOT(d, c) TOF(b, c, a) ]CNOT(d, c) TOF(b, c, d) CNOT(b, c) 
\subsection*{APP2.10}
TOF(a, b, d) CNOT(c, b) TOF(a, b, c) TOF(b, c, d) CNOT(d, c) CNOT(d, b) TOF(b, c, d) CNOT(a, b) TOF(c, d, b) TOF(a, c, d) [CNOT(d, c) TOF(b, c, a) CNOT(d, c) TOF(b, c, a) TOF(b, d, a) ]CNOT(a, d) TOF(c, d, a) TOF(b, d, a) TOF(c, d, a) TOF4(b, c, d, a) 
\subsection*{APP2.11}
TOF(a, c, b) NOT(c) TOF4(b, c, d, a) TOF(a, b, c) CNOT(a, c) TOF4(a, c, d, b) TOF4(a, b, c, d) [CNOT(d, c) CNOT(c, b) TOF(b, c, a) CNOT(c, a) CNOT(c, b) TOF(b, c, a) CNOT(d, c) ]TOF(b, d, a) CNOT(a, c) TOF4(a, c, d, b) CNOT(c, d) TOF(a, b, c) TOF(a, b, d) NOT(c) 
\subsection*{APP2.12}
TOF(c, d, a) TOF(a, c, d) CNOT(c, d) NOT(a) TOF(c, d, a) [CNOT(d, c) CNOT(c, b) TOF(b, c, a) CNOT(c, b) CNOT(d, c) CNOT(c, a) CNOT(d, c) CNOT(c, a) CNOT(d, c) CNOT(c, a) TOF(b, c, a) TOF(b, d, a) ]TOF(a, c, d) NOT(d) CNOT(d, c) TOF(a, c, d) NOT(a) NOT(b) CNOT(a, d) TOF4(a, c, d, b) NOT(c) CNOT(c, a) CNOT(b, c) TOF(a, b, d) 
\subsection*{APP2.13}
CNOT(c, a) TOF(b, d, a) NOT(c) TOF(a, c, d) TOF(a, b, c) TOF4(a, b, d, c) TOF(b, d, a) CNOT(d, a) CNOT(b, d) [CNOT(d, c) TOF(b, d, a) CNOT(d, c) TOF(b, c, a) CNOT(d, c) TOF(b, c, a) CNOT(d, c) ]NOT(a) NOT(d) TOF(a, b, d) TOF(b, c, a) NOT(d) TOF(a, d, b) NOT(c) NOT(d) TOF(b, c, d) 
}

\end{document}